# Observation of giant remnant polarization in ultrathin AlScN at cryogenic temperatures

Seunguk Song[1,2,3,7], Dhiren K. Pradhan[1,7], Zekun Hu[1,7], Yinuo Zhang[1], Rachael N. Keneipp[4], Michael A. Susner[5], Pijush Bhattacharya[5,6] Marija Drndić[4], Roy H. Olsson III[1,*], and Deep Jariwala[1,*]

[1]Department of Electrical and System Engineering, University of Pennsylvania, Philadelphia, Pennsylvania 19104, United States of America
[2]Department of Energy Science, Sungkyunkwan University (SKKU), Suwon, 16419, Republic of Korea
[3]Center for 2D Quantum Heterostructures (2DQH), Institute for Basic Science (IBS), Sungkyunkwan University (SKKU), Suwon, 16419, Republic of Korea
[4]Department of Physics and Astronomy, University of Pennsylvania, Philadelphia, Pennsylvania 19104, United States of America
[5]Materials and Manufacturing Directorate, Air Force Research Laboratory, Wright-Patterson AFB, Dayton, OH, United States of America
[6] Azimuth Corporation, 2079 Presidential Dr. #200, Fairborn, OH 45342, United States of America
[7]These authors equally contributed to this work.
[*]Correspondence should be addressed: rolsson@seas.upenn.edu (R.H.O.) & dmj@seas.upenn.edu (D. J.)

## Abstract

The discovery of wurtzite ferroelectrics opens new frontiers in polar materials, yet their behavior at cryogenic temperatures remains unexplored. Here, we reveal unprecedented ferroelectric properties in ultrathin (10 nm) $Al_{0.68}Sc_{0.32}N$ (AlScN) at cryogenic temperatures where the properties are fundamentally distinct from those of conventional oxide ferroelectrics. At 12 K, we demonstrate a giant remnant polarization exceeding 250 μC/cm$^2$—more than twice that of any known ferroelectric—driven by an enhanced *c/a* ratio in the wurtzite structure. Our devices sustain remarkably high electric fields (~13 MV/cm) while maintaining reliable switching, achieving over $10^4$ polarization reversal cycles at 12 K. Critically, this breakdown field strength approaches that of passive dielectric materials while maintaining ferroelectric functionality. The extraordinary polarization enhancement and high-field stability at cryogenic temperatures contrasts sharply with oxide ferroelectrics, establishing wurtzite ferroelectrics as a distinct class of polar materials with implications spanning fundamental physics to cryogenic non-volatile memory and quantum technologies.

# Introduction

The recent discovery of ferroelectricity in wurtzite-structured $Al_{1-x}Sc_xN$ inaugurates a new class of polar materials fundamentally different from conventional oxide ferroelectrics[1]. While oxide ferroelectrics rely on B-site cation displacement (perovskites) or oxygen vacancy ordering (fluorites), wurtzite ferroelectrics achieve polarization through ionic charge separation in a non-centrosymmetric lattice[2]. This distinct mechanism leads to remarkably high room-temperature polarization (>100 μC/cm$^2$) and enables previously impossible applications from high-temperature memory to ferroelectric diodes[3,4].

The behavior of ferroelectric materials at cryogenic temperatures provides crucial insights into their fundamental physics while enabling emerging cryogenic memory and quantum device technologies[5]. In oxide ferroelectrics, reduced thermal fluctuations at low temperatures typically result in modest polarization changes, with values rarely exceeding 150 μC/cm$^2$ even at 4 K[6-10]. The archetypal $HfO_2$-based systems show only 20-30% polarization enhancement at cryogenic temperatures[11], while some perovskites exhibit decreased polarization[7,9]. This limitation stems from their reliance on structural distortions that become energetically costly at low temperatures.

The electrical properties of materials at cryogenic temperatures are particularly critical for cryo-computing applications in the digital domain and, increasingly, in the analog domain for quantum computing. While conventional dielectric thin films like $SiO_2$ and $Al_2O_3$ show breakdown fields of 13-15 MV/cm[12,13], and emerging 2D dielectrics like h-BN demonstrate ~7.8 MV/cm[14], maintaining such high breakdown fields in functional ferroelectric materials remains elusive. $HfO_2$-based ferroelectrics, for instance, typically breakdown at 7 MV/cm at 4 K[6] and in another report 3.9 MV/cm even at deep cryogenic temperatures (~1.45 K)[15], highlighting a fundamental limitation in oxide ferroelectrics. Moreover, the cryogenic behavior of wurtzite ferroelectrics remains largely unexplored, presenting a critical gap in our understanding of these polar materials. The wurtzite structure's inherent polarity suggests potentially distinct low-temperature physics, as reduced thermal motion can enhance the intrinsic charge separation rather than hinder it.

In this work, we reveal extraordinary enhancement of ferroelectric properties in ultrathin (10 nm) $Al_{0.68}Sc_{0.32}N$ at cryogenic temperatures, achieving simultaneous records in both polarization and breakdown strength. Through temperature-dependent structural and electrical characterization, we demonstrate that cooling to 12 K induces a dramatic increase in the *c/a* ratio, resulting in a giant remnant polarization exceeding 250 μC/cm$^2$—a value unprecedented in any known ferroelectric material at any temperatures. Remarkably, our devices sustain electric fields approaching ~13 MV/cm without breakdown at 12 K, nearing the performance of the best passive dielectric materials while maintaining reliable ferroelectric switching. This combination of ultra-high polarization and breakdown strength enables enhanced endurance (>10$^4$ cycles) at cryogenic temperatures compared to the performance at room temperature. This exceptional behavior establishes wurtzite ferroelectrics as a distinct class of polar materials that is governed by physical principles fundamentally different from oxide ferroelectrics. This intrinsic quality has immediate implications for both cryogenic computing and our understanding of polar material physics.

## Results and Discussion

$Al_{1-x}Sc_xN$ (AlScN) thin films with a scandium concentration of $x = 0.32$ and thicknesses of 10 or 30 nm were grown on Al(111)/sapphire wafers (**Fig. 1a**) by reactive pulsed DC magnetron co-sputtering (Methods). To promote high-quality growth, the $N_2$ flow was set to a higher level (~27.5 sccm) than in our previous report for thicker AlScN (~20 sccm)[16]. We then performed X-ray diffraction (XRD) measurements using Co K-alpha radiation at different temperatures from ~80 to 300 K to investigate the crystal structures (**Fig. 1d-k**). For both the 10 and 30 nm-thick AlScN films, the XRD data show peaks corresponding to Al(111), $Al_2O_3$(006), and a *c*-axis-oriented AlScN(002) (**Fig. 1d,g**). To highlight the temperature-dependent structural behavior of the 10-nm-thick AlScN, we present zoomed-in XRD data for the AlScN(002) and Al(111) peaks in **Fig. 1e** and **1f**, respectively. As the temperature decreases from RT to 100 K, the *d*-spacing of AlScN(002) increases from ~2.49 Å to ~2.50 Å (**Fig. 1e**). In contrast the *d*-spacing of Al(111) decreases from ~2.335 Å to ~2.320 Å near 100 K (**Fig. 1f**), indicating compressive strain along the *a*-axis of AlScN and lattice anisotropy at lower temperatures. Such compressive strain may lead to a pronounced expansion of the ratio between lattice constants *c* and *a* (i.e., increased *c*/*a* ratio) in the hexagonal wurtzite structure of AlScN at lower temperatures, a phenomenon also referred to as negative thermal expansion (**Fig. 1b,c**).

Compared with the 10 nm film, the 30 nm AlScN (**Fig. 1g-i**) exhibits less anisotropic strain. The *d*-spacing of AlScN(002) in the 30 nm film remains nearly constant (~2.485-2.490 Å) over the measured temperatures (~80-300 K) (**Fig. 1h**), suggesting a weaker negative thermal expansion. The more pronounced structural changes in the thinner film (10 nm) are likely attributed to the higher strain induced by its reduced thickness, which may in turn affect its ferroelectric behavior at low temperatures. For example, a higher *c*/*a* ratio of AlScN is closely associated with greater anisotropy in the wurtzite structure, increased polarity of the Al-N bonds, and an elevated ferroelectric switching energy barrier[17-20].

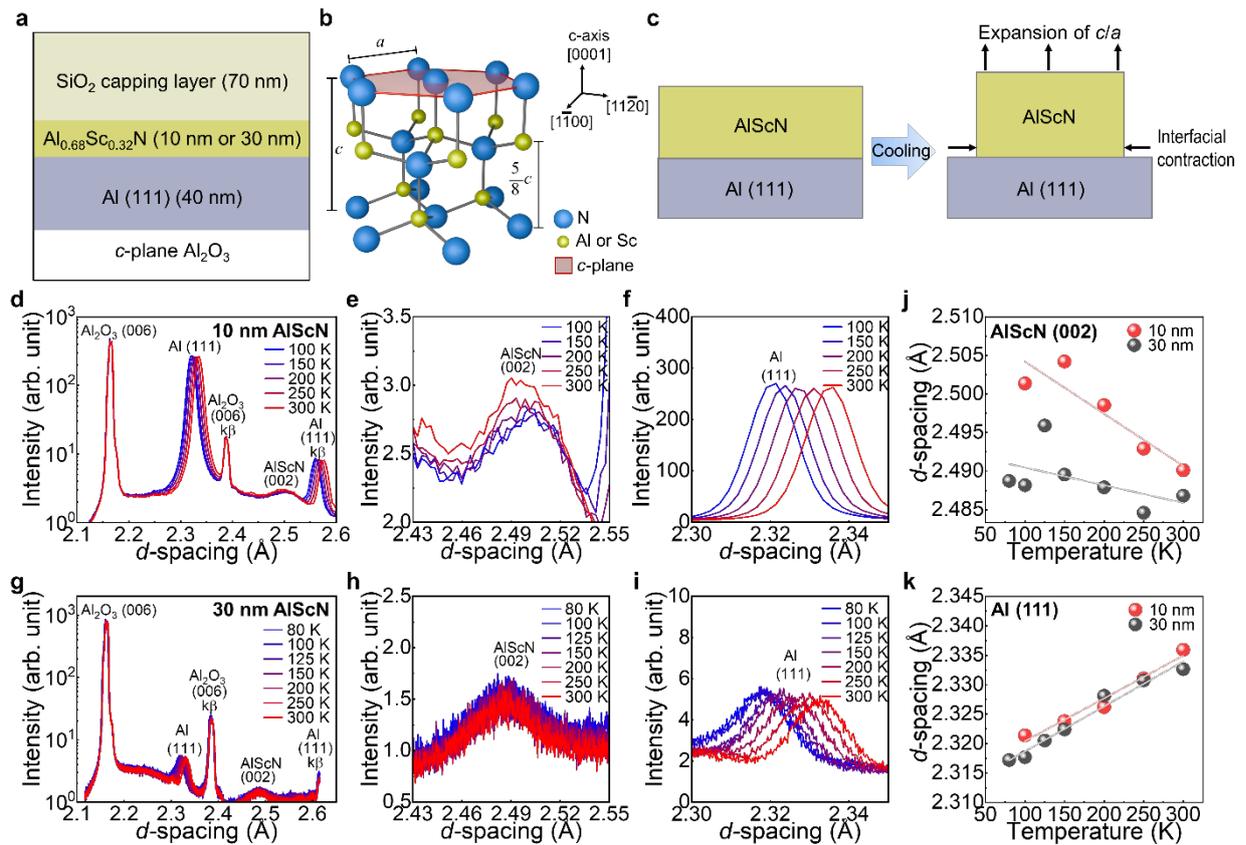

**Figure 1. Structural analysis of $Al_{1-x}Sc_xN$ (x = 0.32) thin films grown at high $N_2$ ambient.** (**a**) Cross-sectional schematic illustration of the 10 and 30 nm-thick AlScN on Al(111)/sapphire wafers. (**b, c**) Schematics depicting the temperature-dependent variations in lattice constants *a* and *c* of AlScN. (**d-i**) X-ray diffraction (XRD) patterns collected at different temperatures (∼80–300 K), showing peaks corresponding to Al(111), $Al_2O_3$(006), and AlScN(002) of the AlScN with the thickness of (**d-f**) 10 nm and (**g-i**) 30 nm. (**e,f,h,i**) Zoomed-in views of the XRD (**e,h**) AlScN(002) and (**f,i**) Al(111) *d*-spacings for AlScN films with (**e,f**) 10 nm and (**h,i**) 30 nm thicknesses. (**j, k**) comparisons of *d*-spacing of 10 nm and 30 nm AlScN films for (**j**) AlScN(002) and (**k**) Al(001) planes. Solid lines are guides to eye to highlight the trends.

A 3D schematic depiction of a via-connected ferroelectric capacitor structure with 10 nm of $Al_{0.68}Sc_{0.32}N$ (AlScN) is shown in **Fig. 2a**, and **Fig. 2b** provides a cross-sectional view of AlScN and other layers. **Figure 2c** displays an optical microscopic (OM) image of the top surface of the devices, featuring via-connected circular top Al electrodes over the AlScN ferroelectric layer. The use of vias and contact pads, instead of a directly connected top electrode, prevents unnecessary mechanical strain on the ultrathin ferroelectric capacitors during probing. The top probing pad has a diameter of 50 μm, while the via is 4 μm in diameter, positioned on a 10 μm diameter capacitor contact. We chose circular shapes for the top Al electrodes to ensure a homogeneous electric field and to minimize edge effects.

For these ferroelectric capacitors, we performed triangle wave current density vs. electric field (*J–E*) measurements over a wide temperature range from room temperature to 12 K in 50 K increments. We applied a 25 kHz triangular pulse to the devices and the excitation voltage gradually increased until device failure. Triangle wave *I–V* measurements offer a dynamic approach to probe ferroelectric switching at higher voltage ramp rates, thereby distinguishing ferroelectric switching currents from leakage.

**Figure 2d-g** presents the triangle-wave *J-E* behavior of AlScN-based ferroelectric capacitors measured at 300, 200, 100 and 12 K, respectively (see **Fig. S1a-d** for data at 50, 150 and 250 K). Ferroelectric switching is observed at all temperatures down to 12 K, confirming robust ferroelectricity even at this low temperature. A prominent peak in the switching current defines the switching voltage in the *J-E* curves, from which the coercive field ($E_C$) is derived. For instance, **Fig. 2h** (and **Fig. S1e-h**) illustrates how $E_C$ was extracted at various temperatures using the logarithm plot. The red and black arrows indicate the $E_C$ of 12 and 300 K, respectively. Notably, these switching peaks become more distinguishable at lower temperatures (**Fig. 2e-f**) compared with RT (**Fig. 2d**), due to the suppression of the thermionic-assisted leakage currents.

**Figure 2i** shows the temperature dependence of $E_C$ and breakdown field ($E_B$) down to 12 K. Across all temperatures, $E_B$ remains higher than $E_C$, and both increase with decreasing temperature. This behavior indicates that AlScN devices tolerate higher electric fields at low temperatures, presumably because of reduced ion movement that otherwise facilitates early breakdown at higher temperatures. Additionally, the crystal structure at lower temperatures further enhances the breakdown field, resulting in a higher $E_B$ of > ~13.5 MV/cm at 12 K, compared with ~11.8 MV/ at RT. In **Fig. 2j**, the ratio $E_B/E_C$ is comparted for different N$_2$ flow conditions during AlScN deposition (See also *J-E* curves of AlScN with low N$_2$ flow[16] in **Fig S2.**). The $E_B/E_C$ is found to be increasing over temperatures in both films deposited under higher and lower N$_2$ flow. However, the AlScN films grown under higher N$_2$ flow exhibit higher $E_B/E_C$. We hypothesize that higher N$_2$ flow reduces concentration of nitrogen vacancies, which enhances the structural stability of the material and resistance to breakdown under high electric fields, which enables characterizations of ferroelectric switching down to 12 K.

**Figure 2k** plots the positive coercive field ($E_C^+$) and negative coercive field ($E_C^-$) as a function of temperature down to 12 K, derived from the triangle-wave *J-E* measurements. Both $E_C^+$ and $E_C^-$ increase linearly as the temperature decreases. The $E_C^+$ values at 12 K and at RT are observed to be ~6.4 and ~10 (+/- 0.1) MV/cm, respectively, whereas the $E_C^-$ values at these temperatures are found to be ~-6.2 and ~-9.8 (+/- 0.1) MV/cm, respectively. At RT, the values of $E_C^+$ and $E_C^-$ for similar thicknesses are well matched with the earlier reports[21]. This increase in $E_C$ with lower temperatures can be attributed to the reduced mobility of domain walls, which heightens the energy required to switch the domains, thereby raising the coercive field[7]. Generally, coercive field is seen as the threshold field at which the domain walls overcome pinning forces, requiring higher fields at lower temperatures due to reduced thermal activation. A similar trend has been reported in PZT capacitors[7,22]. Moreover, in a study on domain-wall propagation in epitaxial PZT, the increase of $E_C$ was linked to temperature-dependent dynamics of domain wall motion, which transitions from thermally activated creep at higher temperatures to being dominated by pinning effects from quenched defects at lower temperatures[10].

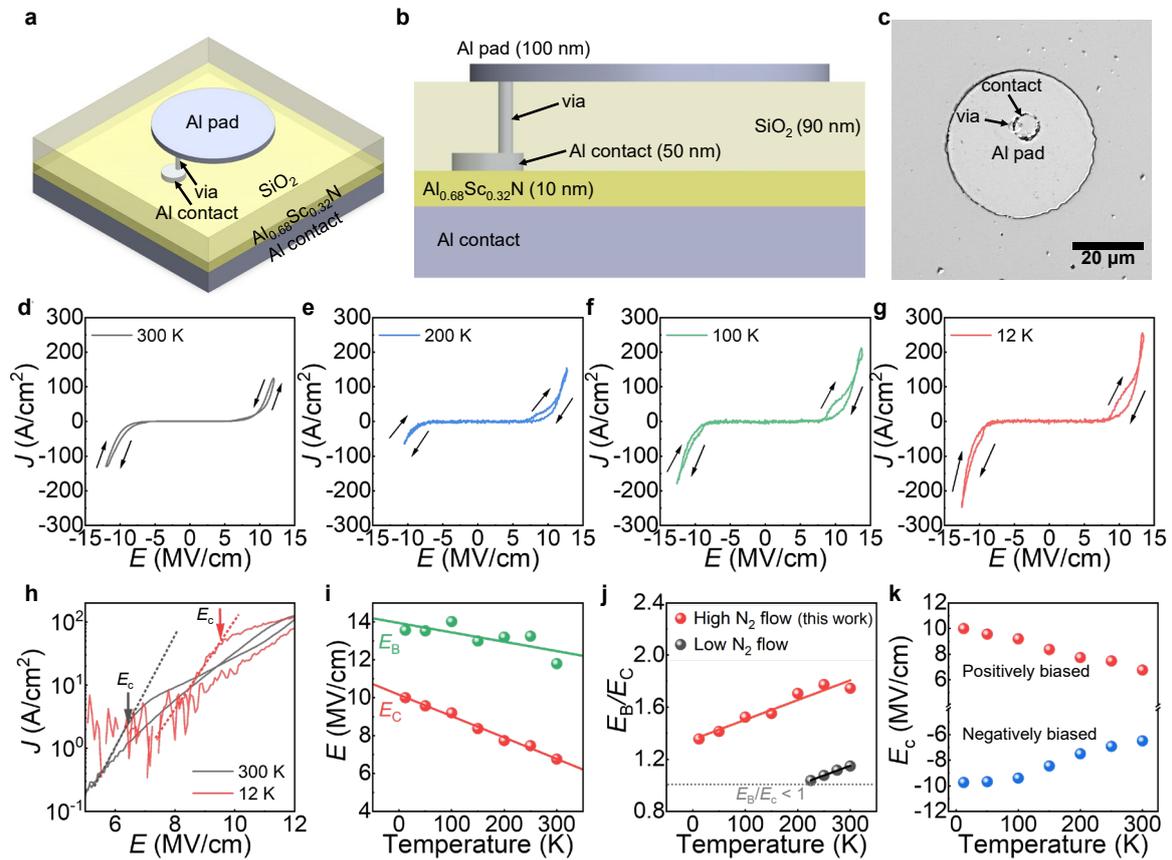

**Figure 2. Temperature-dependent ferroelectric switching of Al$_{0.68}$Sc$_{0.32}$N capacitors with via structure.** (**a**) 3D schematic of via-connected ferroelectric capacitor (FeCap) structure; (**b**) cross-sectional view showing Al$_{0.68}$Sc$_{0.32}$N and other layers; (**c**) optical image of the FeCap device. (**d–g**) Triangle wave electric field (*E*) vs. current density (*J*) hysteresis loops at 300, 200, 100, and 12 K, showing temperature-dependent ferroelectric switching. (**h**) Comparison of coercive field $E_C$ at 300 K and 12 K, and the extraction of $E_c$ is labeled. (**i**) Breakdown field ($E_B$) and coercive field ($E_C$) as a function of temperatures, with decrease of temperatures. (**j**) Temperature-dependent ratio of $E_B$ to $E_C$ of AlScN

capacitance with different N$_2$ flow conditions, indicating the impact on ferroelectric stability. (**k**) Positive and negative coercive field values as a function of temperature, showing symmetric behavior. Solid lines in (i) and (j) are guides to eye to highlight the trends.

To further investigate ferroelectric properties and polarization dynamics of the 10-nm-thick AlScN ferroelectric capacitors, Positive Up Negative Down (PUND) measurements were performed with square pulses with rise/fall times of 20 ns and a pulse width of 0.5 μs over a temperature range of ~12 to 300 K. The PUND technique enables accurate determination of the remanent polarization ($P_r$) by isolating the pure ferroelectric response from leakage and other non-ferroelectric currents. **Figure 3a-c** present the measured currents under an applied bias of 13 V at 12 K, and **Fig. 3b** shows the current response for the "P-U" sequences, where a distinct current peak corresponding to ferroelectric switching (labeled "FE") is observed during the "P" pulse. Similarly, ferroelectric switching is evident under the "N" pulse during the "N-D" sequence (**Fig. 3c**). This clearly indicates that ferroelectric switching occurs even at the low temperature of 13 K.

**Figure 3d** (and **Fig. S3**) displays the current responses measured at the same applied voltage of ~13 V at different temperatures (13-300 K). The corresponding "P-U" and "N-D" responses are shown in **Fig. 3e** and **3f**, respectively. Although the ferroelectric current contributions (i.e., the "FE" peaks) remain nearly the same, the leakage components diminish more as the temperature decreases (indicated as arrows). Because the switching polarization ($2P_r$) is determined by subtracting the polarization measured during the "U" and "D" pulses from that measured during the "P" and "N" pulses, the temperature dependence of these current responses implies that the $2P_r$ increases at lower temperature down to 12 K for the same applied voltage.

In contrast to the PUND results of the 10 nm AlScN film (**Fig. 3a-f**), in the 30 nm AlScN film (**Fig. 3g-i**), the FE current gradually decreases as the temperature is lowered under the same electrical field of ~7.5 MV/cm. Below 200 K, the observation of FE signals gets challenging under the same field, probably because of the leakage current dominating the curve for the thicker film (see also **Fig. S4** for more studies on the 30 nm AlScN). Similarly, for 45 nm AlScN deposited with low N$_2$ flow, the ferroelectric current intensity decreases as the temperature lowers (**Fig. S5**); however, the PUND measurement is impossible below 200 K to see the FE currents even with a high electrical field due to the premature electrical breakdown.

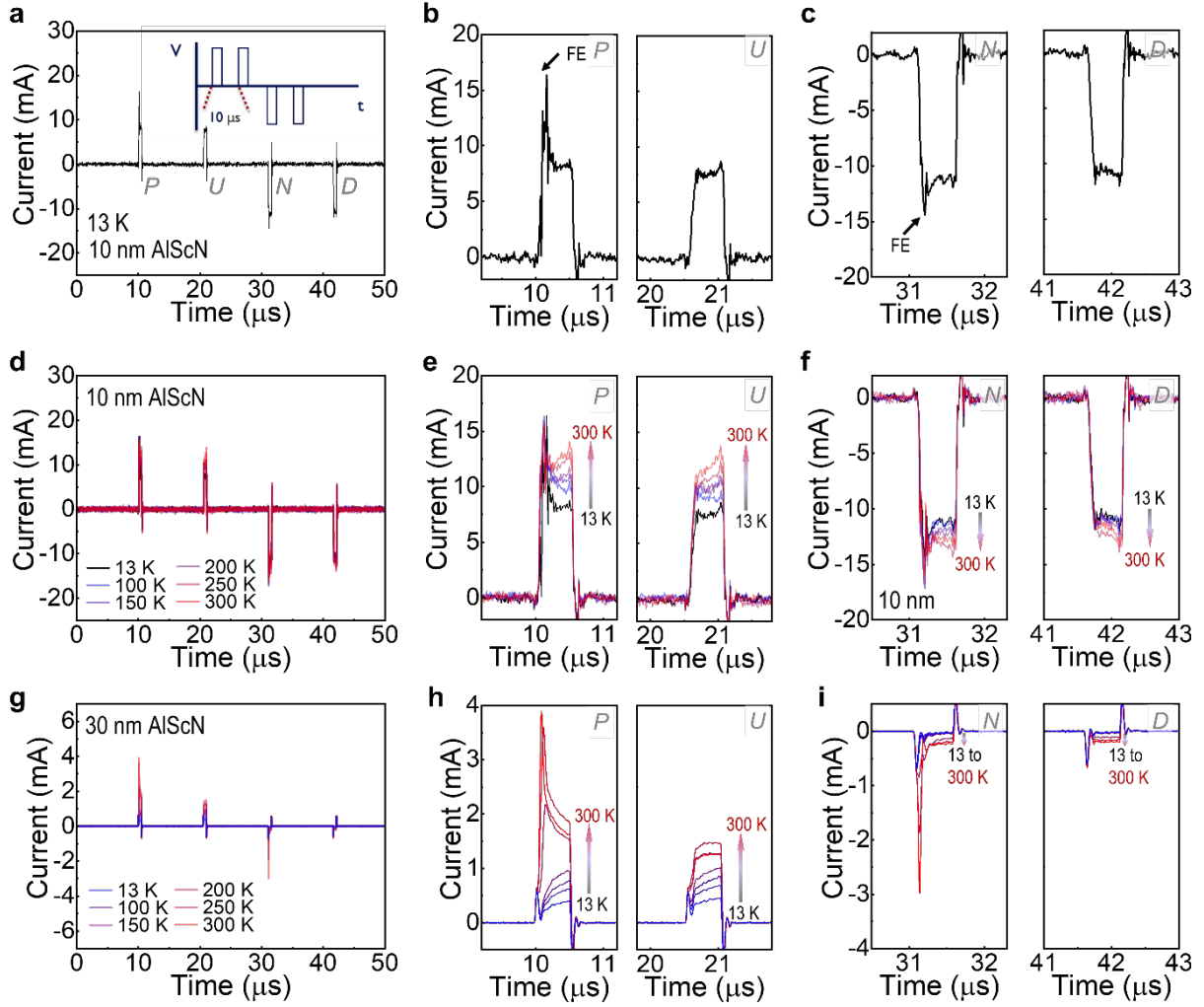

**Figure 3. Temperature-dependent ferroelectric properties and polarization switching of $Al_{0.68}Sc_{0.32}N$ capacitors with via-contacted structure.** (**a**) Positive-Up Negative-Down (PUND) current trace of 10 nm AlScN ferroelectric capacitance at 13 K. Inset shows the voltage pulse applied during PUND measurement, (**b**) P-U and (**c**) N-D traces at 13 K indicating the conclusive evidence of ferroelectric polarization switching in both positive and negative sides. (**d-f**) PUND traces of 10-nm-thick AlScN at various temperatures of 13-300 K, and the zoomed-in (**e**) P-U (**f**) N-D traces. (**g-i**) PUND measurement results of 30-nm-thick AlScN at various temperatures. All the PUND currents here are recorded as a response of a same pulse amplitude of 7.5 MV/cm.

By integrating the ferroelectric and resistive components of the current, we estimate $2P_r$ of 10 nm AlScN as a function of the applied field ($E$) at different temperatures (**Fig. 4a**). As temperature increases, the $2P_r$ values start rising at smaller $E$, reflecting a smaller coercive field ($E_C$) required to switch the polarization (See also a comparison of $E_c$ extracted by hysteresis and PUND in **Fig. S6a.**). Likewise, the $2P_r$ saturates at lower $E$ values for higher temperatures because the smaller $E_c$ allows complete ferroelectric switching of the capacitor under a reduced field (The extraction of the saturated $2P_r$ is depicted in **Fig. S6b.**).

**Figure 4b** summarizes the saturated $P_r$ in AlScN capacitors with 10 nm (solid) and 30 nm (open) thicknesses at different temperatures. Although a slight asymmetry exists between the responses of positively biased and negatively biased pulses, the overall trend shows that $P_r$ values increase from ~173-224 μC/cm$^2$ to ~272-359 μC/cm$^2$ as the temperature decreases from RT to 13 K for the 10 nm AlScN. On the other hand, the 30 nm films show a similar $P_r$ of ~56-66 μC/cm$^2$ in the temperature range of 200-300 K (See also **Fig. S4** for more studies on 30 nm thick film.). We suggest that, for our ~10 nm-thick AlScN film, cooling leads to an increase in the $c/a$ ratio and polarity of the Al-N bonds caused by strain effects at the Al/AlScN interface (**Fig. 1**). This enhanced $c/a$ ratio of AlScN at low temperatures might give rise to a remarkably high $P_r$ of ~272–359 μC/cm$^2$ for the 10-nm-thick film. The enhancement of polarization at lower temperatures might also be attributed to the reduction in thermal agitation, which stabilizes dipole alignment and restricts domain wall motion, thereby facilitating enhanced ferroelectric ordering[6,23].

Our AlScN exhibits significantly enhanced $P_r$ at low temperatures, surpassing the values reported for other materials, including Zr- or Si-doped HfO$_2$ (HZO)[6,11,24], PZT[7,25], and SrBi$_2$Ta$_2$O$_9$ (SBT)[9] (**Fig. 4c** and **Table S1**) While previous work on 10 nm ALD-grown HZO reported a $2P_r$ of up to 150 μC/cm$^2$ at 4 K[6], our sputtered 10 nm AlScN film achieves a significantly higher $2P_r$ of > ~544 μC/cm$^2$ at 13 K (negatively biased side). Notably, while some conventional ferroelectrics (e.g., SBT)[9] experience a significant drop in $2P_r$ upon cooling, AlScN maintains and even enhances its $2P_r$ at cryogenic temperatures. This exceptional behavior highlights the importance of AlScN interfaces in achieving high $2P_r$, which is critical for low-temperature device performance. However, despite AlScN's remarkably high $2P_r$ at low temperatures, its $E_C$—although below 10 MV/cm—remains notably higher than that of other ferroelectric thin films such as HZO[6,11,24] or PZT[7,25], which typically exhibit $E_C$ values in the range of 0.7 to 3 MV/cm (**Fig. 4d**). Although, optimizing AlScN to achieve a lower $E_C$ while preserving its high $2P_r$ would be beneficial for developing low-power cryogenic ferroelectric devices for the future, ultrathin AlScN possesses ultrahigh $E_C$, $E_B$ and $P_r$ at cryogenic temperatures which is unique for a functional dielectric material (see the comparisons in **Fig. 4c-f** and **Table S1**).

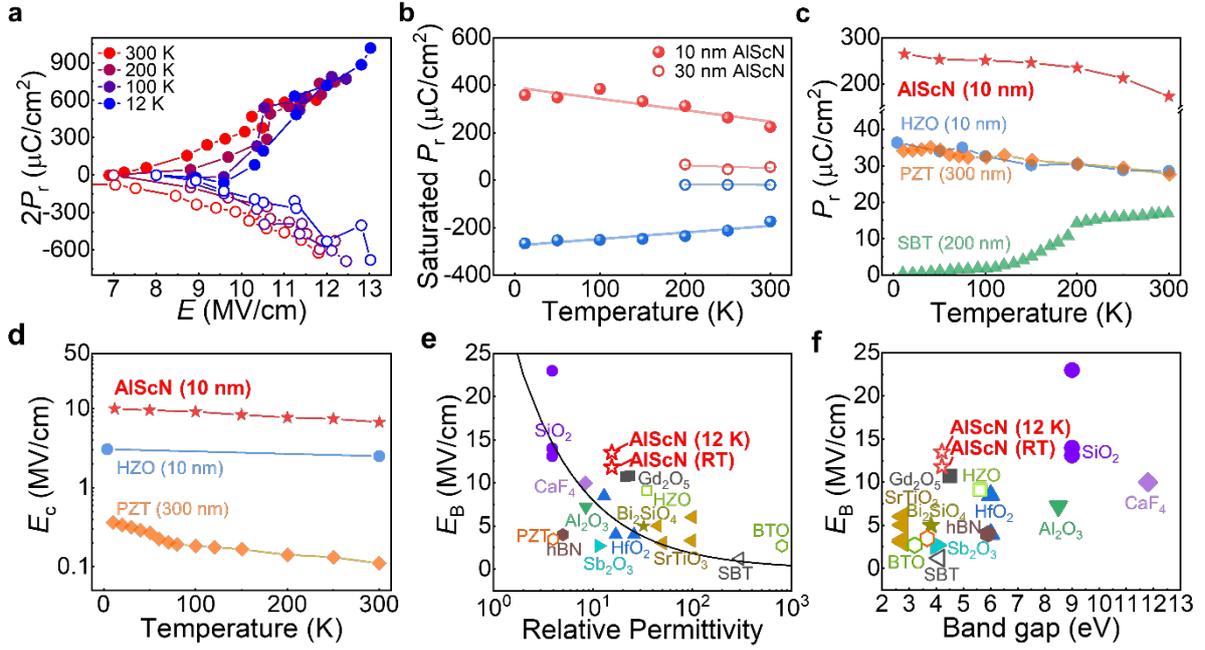

**Figure 4. Temperature-dependent ferroelectric polarization and coercivity in AlScN.** (**a**) Remnant polarization ($2P_r$) as a function of the PUND electric field at 12, 100, 200, and 300 K. (**b**) Temperature dependence of the saturated $P_r$, extracted from the PUND measurements under positively (red) and negatively (blue) biased fields. 10 nm and 30 nm AlScN results are displayed in solid and open circles, respectively. Solid lines in (b) are guides to eye to show the trends. (**c,d**) Comparison of (**c**) remnant polarization ($P_r$) and (**d**) coercive field ($E_C$) for our AlScN (red; negatively biased ones from (b)) and other various ferroelectric materials (e.g., HZO[6], PZT[7], and SBT[9]) as a function of temperature. (**e,f**) Breakdown field ($E_B$) versus (**e**) relative permittivity, and (**f**) bandgap of typical insulators (i.e., $Gd_2O_5$[26], $SiO_2$[26], $HfO_2$[27,28], $Al_2O_3$[29], $CaF_4$[30], $SrTiO_3$[31], $Sb_2O_3$[32], $hBN$[33], $Bi_2SiO_5$[34]; solid symbols) and ferroelectrics (i.e., PZT[35], BTO[36], SBT[37], HZO[27,38]; open symbols) at RT.

We further conducted write-cycling endurance tests of via-contacted 10 nm AlScN ferroelectric capacitors at different temperatures using symmetric PUND pulses. The pulse parameters matched those used in the PUND measurements shown in **Fig. 3**. For each temperature tested, we applied pulse amplitudes exceeding $E_c$ (e.g., $E_c \approx 10$ MV/cm at 12 K in **Fig. 2k**) to ensure complete polarization switching and extraction of P-U and N-D charges (**Fig. S7**). The endurance characteristics at RT and 12 K under various applied fields are presented in **Fig. 5a** and **5b**, respectively. When the applied field ($E$) exceeds the $E_c$ at each temperature, the capacitors demonstrate endurances beyond $10^4$ cycles. Notably, low-temperature operation significantly improves endurance performance under identical applied fields. For instance, at an $E$ of ~10.5 MV/cm, the device sustains over $>10^3$ cycles at RT and exceeds $>10^4$ cycles at 12 K.

**Figure 5c** and **5d** illustrate how cycle counts vary with temperatures as functions of applied $E$ and extracted $2P_r$ values. The data consistently shows superior endurance characteristics at 12 K compared to RT, regardless of the applied $E$ or $2P_r$ values. This enhancement at lower temperatures can be attributed to reduced electron tunneling, decreased vacancy-related conduction, and diminished heating effects. Specifically, nitrogen vacancies near the interface can cause downward band bending in AlScN, leading to increased tunneling currents—an effect that becomes more pronounced at higher temperatures due to additional thermal energy and higher leakage currents (as indicated in **Fig. 5d-f**). The reduced temperature minimizes device degradation from vacancy-induced leakage and Joule heating, resulting in higher endurance cycles. This improvement enables the ferroelectric capacitors to sustain more than $10^4$ cycles at 12 K even with high $2P_r$ values exceeding > 400 μC/cm$^2$ (achieved at $E$ > ~12 MV/cm; from negative N-U pulses).

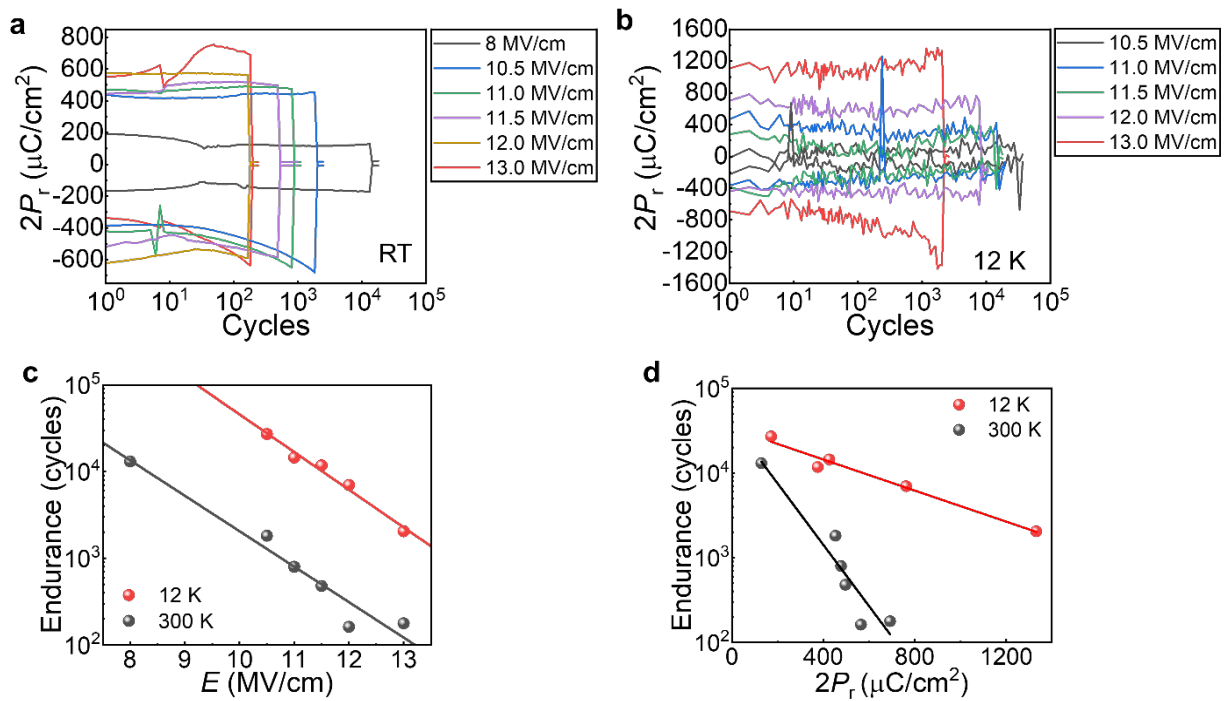

**Figure 5. Temperature-dependent PUND endurance of AlScN capacitors with via structure.** PUND endurance write cycles at different applied electric fields at (**a**) 300 K and (**b**) 12 K. (**c**) PUND endurance cycle as a function of applied electric field at 12 and 300 K. (**d**) PUND endurance cycle as a function of $2P_r$ at 12 and 300 K. Solid line fitting in shown (c) and (d) are guides to the eyes.

## Conclusions

We have studied the cryogenic ferroelectric behavior of 10-nm-thick $Al_{0.68}Sc_{0.32}N$ (AlScN) films grown at high $N_2$ ambient. X-ray diffraction studies show that the thin film experiences significant compression, reflected in its expanded *c/a* ratio at lower temperatures, which enhance the polarization at low temperatures. The PUND measurements of the via-contacted AlScN capacitors prove reliable ferroelectric switching at different low temperatures and reveals remarkably high saturated $P_r$ values over 250 μC/cm² at 12 K. The increased breakdown field ($E_B$) of AlScN grown with higher $N_2$ flow leads to better $E_B/E_c$ ratios, leading to reliable operation even under high electrical field. Additionally, endurance measurements reveal that the devices could sustain > $10^4$ switching cycles at 12 K under fields exceeding 10 MV/cm, significantly surpassing their room-temperature performance. This improved ferroelectric polarizations and endurance at lower temperatures paves the way for highly reliable cryogenic ferroelectric applications of AlScN.

## Methods

### Growth of $Al_{0.68}Sc_{0.32}N$

The AlScN thin films of 10 and 30 nm with 32% Sc concentration ($Al_{0.68}Sc_{0.32}N$) were grown on (111)- oriented Al (50 nm) on c-axis oriented sapphire wafers using a physical vapor deposition (PVD) system. First, a 50 nm thick of Al was deposited on the c-axis oriented sapphire wafer by dc-sputtering a temperature of 150 ºC with an Ar gas flow of 20 sccm. Without breaking vacuum, AlScN films were grown from separate Al and Sc targets. The co-sputtering was carried out in an Evatec CLUSTERLINEVR 200 II pulsed DC PVD system with a power of 1000 W/cm² for Al and 655 W/cm² for Sc. Both targets are 100 mm in diameter. The deposition took place at a temperature of 350 ºC under a vacuum of 8 x $10^{-4}$ torr, with $N_2$ flow of 27.5 sccm. Followed by the deposition of the AlScN film, a 50 nm thick Al capping layer was deposited on the top of it. The Al capping layer was deposited in situ without breaking the vacuum, effectively preventing surface oxidation of the AlScN film.

### Device Fabrication

The in-situ aluminum top surface was patterned using a photomask and S1813 resist in conjunction with a mask aligner (MA6 Mask Aligner). Following this, an inductively coupled plasma reactive ion etching (ICP RIE) process (Oxford Cobra ICP) was performed using a $BCl_3/Cl_2$ gas mixture to etch the aluminum. The sample was then immersed in KL remover for 1.5 hours with ultrasonic agitation to ensure complete removal of the resist. After patterning the in-situ aluminum, a 90 nm layer of $SiO_2$ was deposited at 200°C via plasma-enhanced chemical vapor deposition (PECVD) (Oxford PlasmaLab 100). A second photolithography step was then performed using S1813 resist to define the via pattern. To etch the $SiO_2$ in the via, a reactive ion etching (RIE) process (Oxford 80 Plus RIE) was carried out using $CF_4$ gas. The sample was subsequently immersed in KL remover for 1.5 hours at 65°C with ultrasonic agitation to ensure complete resist removal. Finally, a 150 nm layer of aluminum was deposited using physical vapor deposition (Explorer14 Sputterer), followed by a third photolithography process using S1813 resist. The aluminum was then etched using an ICP RIE process with a $Cl_2/BCl_3$ gas mixture. After etching, the sample was immersed in KL remover for 1.5 hours at 65°C with ultrasonic agitation to remove the resist completely.

**Cryogenic XRD characterizations**

We performed room temperature powder diffraction of the thin film material using a Bruker D8 Discover DaVinci system using Co kα radiation (λ = 1.78897 Å). We took variable temperature measurements (down to 70 K base temperature), under vacuum, using an Oxford Chimera stage. Prior to each actual measurement, we waited at least 10 minutes for the temperature to stabilize.

**Device Characterizations**

Electrical characterizations, including *J-E* hysteresis, PUND, retention, and endurance tests, were performed using a Keithley 4200A-SCS parameter analyzer equipped with a NVM library. The low temperature measurements were conducted in a closed-cycle cryogenic probe station (Advanced Research Systems, Inc.). For *J-E* hysteresis analysis, a bipolar triangular waveform at 25 kHz was applied to investigate the temperature dependence of the coercive field. In PUND and endurance assessments, each pulse in the sequence consisted of a voltage pulse with a 20 ns rise/fall time and a 0.5 μs pulse width. All electrical measurements were conducted on top Al circular electrodes with a 10 μm diameter via-structure unless otherwise specified, with electrodes positioned on $Al_{0.68}Sc_{0.32}N$/Al samples. Maximum PUND, *I-V*, and write voltages were varied with changing temperature to account for temperature-variable switching voltage.

## Author contributions

D.J., R.H.O., S.S., D.P., and Z.H. conceived the devices and measurements idea/concepts. Y.Z. deposited the AlScN films, fabricated the via-contact devices and verified room temperature ferroelectricity of the films and devices. S.S., D.P., and Z.H. measured the devices at various temperatures with assistance from R.N.K. and M.D. M.A.S., and P.B. measured the XRD patterns.

## Data availability

All data are available in the paper and Supplementary Information are available from the corresponding authors upon reasonable request.

Supplementary information for

# Observation of giant remnant polarization in ultrathin AlScN at cryogenic temperatures

Seunguk Song[1,2,3,7], Dhiren K. Pradhan[1,6], Zekun Hu[1,7], Yinuo Zhang[1], Rachael N. Keneipp[4], Michael A. Susner[5], Pijush Bhattacharya[5,6] Marija Drndić[4], Roy H. Olsson III[1,*], and Deep Jariwala[1,*]

[1]*Department of Electrical and System Engineering, University of Pennsylvania, Philadelphia, Pennsylvania 19104, United States of America*
[2]*Department of Energy Science, Sungkyunkwan University (SKKU), Suwon, 16419, Republic of Korea*
[3]*Center for 2D Quantum Heterostructures (2DQH), Institute for Basic Science (IBS), Sungkyunkwan University (SKKU), Suwon, 16419, Republic of Korea*
[4]*Department of Physics and Astronomy, University of Pennsylvania, Philadelphia, Pennsylvania 19104, United States of America*
[5]*Materials and Manufacturing Directorate, Air Force Research Laboratory, Wright-Patterson AFB, Dayton, OH, United States of America*
[6] *Azimuth Corporation, 2079 Presidential Dr. #200, Fairborn, OH 45342, United States of America*
[7]*These authors equally contributed to this work.*
[*]Correspondence should be addressed: rolsson@seas.upenn.edu (R.H.O.) & dmj@seas.upenn.edu (D. J.)

**Figure S1-S7**

**Table S1**

**Supplementary References**

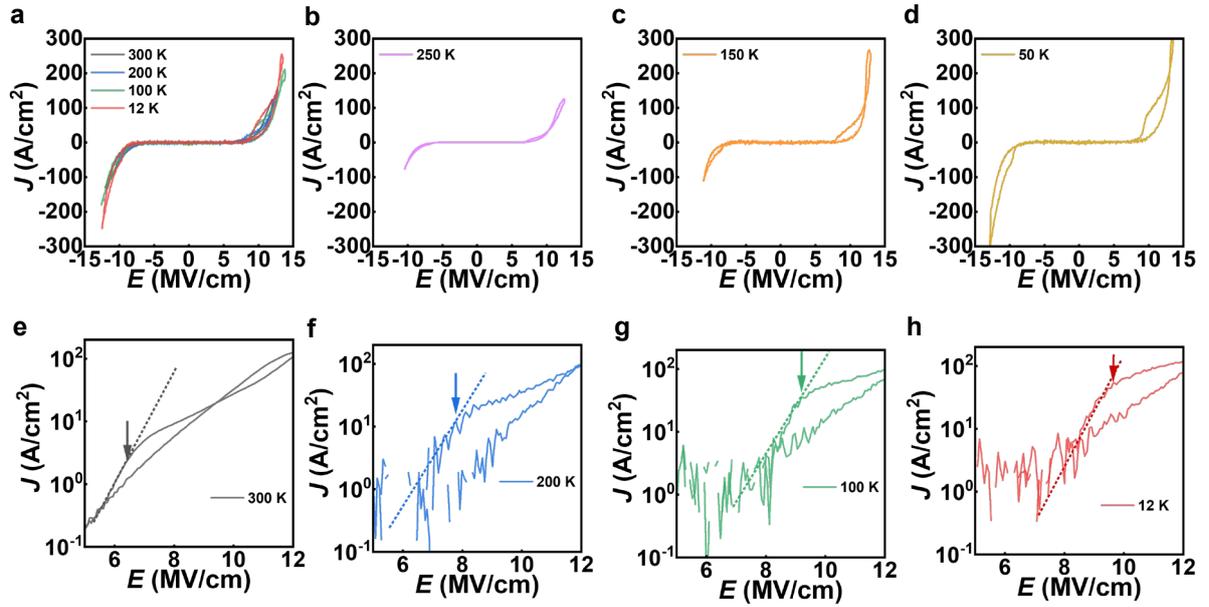

**Figure S1. *J-E* hysteresis curve of 10-nm-thick AlScN capacitance with via contact.** (**a-d**) Triangle-wave *J-E* behavior of AlScN-based ferroelectric capacitors measured at 50, 150, and 250 K, complementing the data at 300, 200, 100, and 12 K presented in Figure 2d-g. The switching peaks in the *J-E* curves become more pronounced at lower temperatures, which can be attributed to the suppression of thermionic-assisted leakage currents. (**e-h**) Logarithmic plots illustrating how the coercive field ($E_c$) is extracted from the switching voltage. The arrows mark the slopes corresponding to $E_c$.

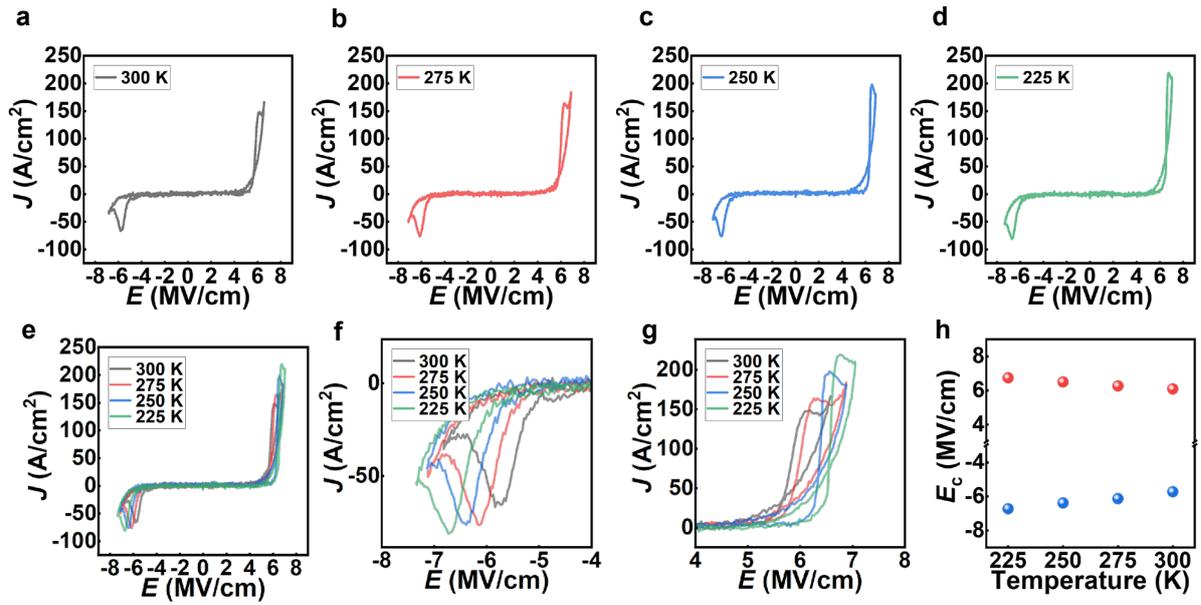

**Figure S2. Hysteresis Analysis of 45 nm AlScN with low $N_2$ flow.** (**a-g**) *J-E* hysteresis of the AlScN capacitors at 225-300K. (**h**) Extracted $E_c$ depending on the temperature. The $E_B$ was ~7 MV/cm regardless of the temperature. Hence, below 225 K, we cannot observe ferroelectric switching because of the failure at the low temperature (i.e., $E_B/E_c < 1$).

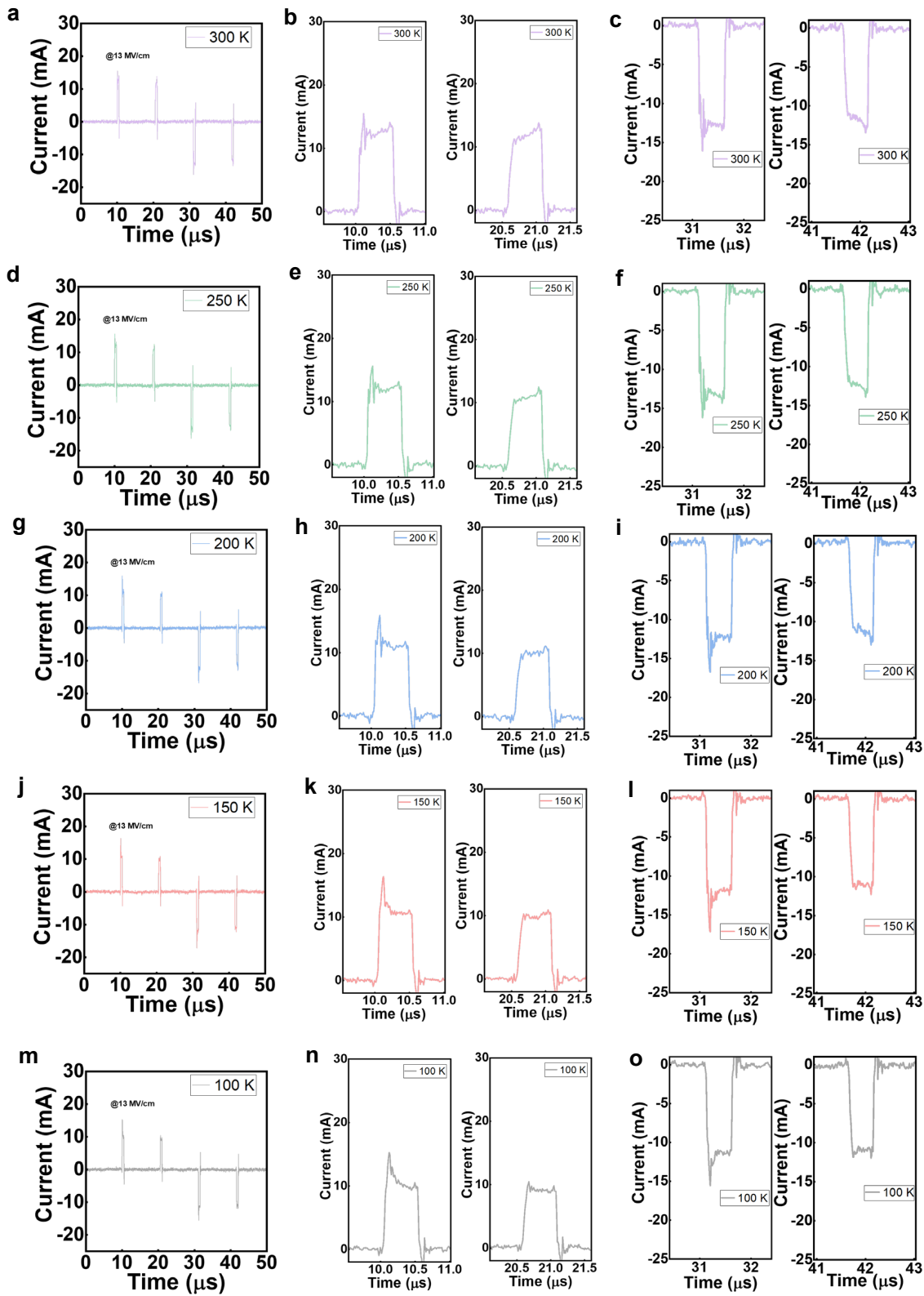

**Figure S3. PUND currents of 10-nm-thick AlScN capacitance with via contact.** Current responses measured at the same applied voltage of ~13 V over a temperature range of 13–300 K, identical to those shown in Figure 3d. Here, the data are separated to reduce complexity caused by overlapping traces.

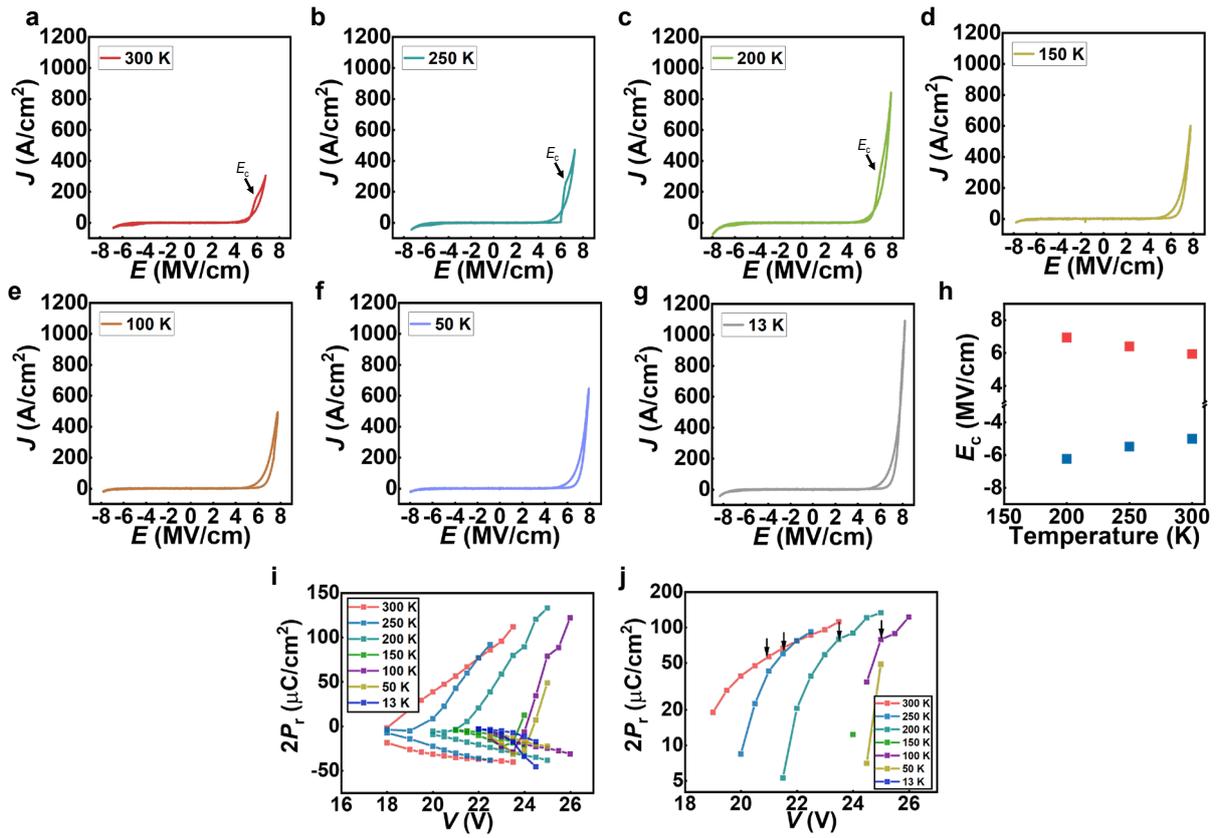

**Figure S4. 30 nm AlScN capacitance with via-connected devices.** (**a-g**) J-E hysteresis curves at different temperatures, and (**h**) extracted coercive fields ($E_c$). (**i, j**) PUND-derived $2P_r$ values depending on the applied voltage (V). Arrows in (**j**) indicate the saturated $2P_r$ values.

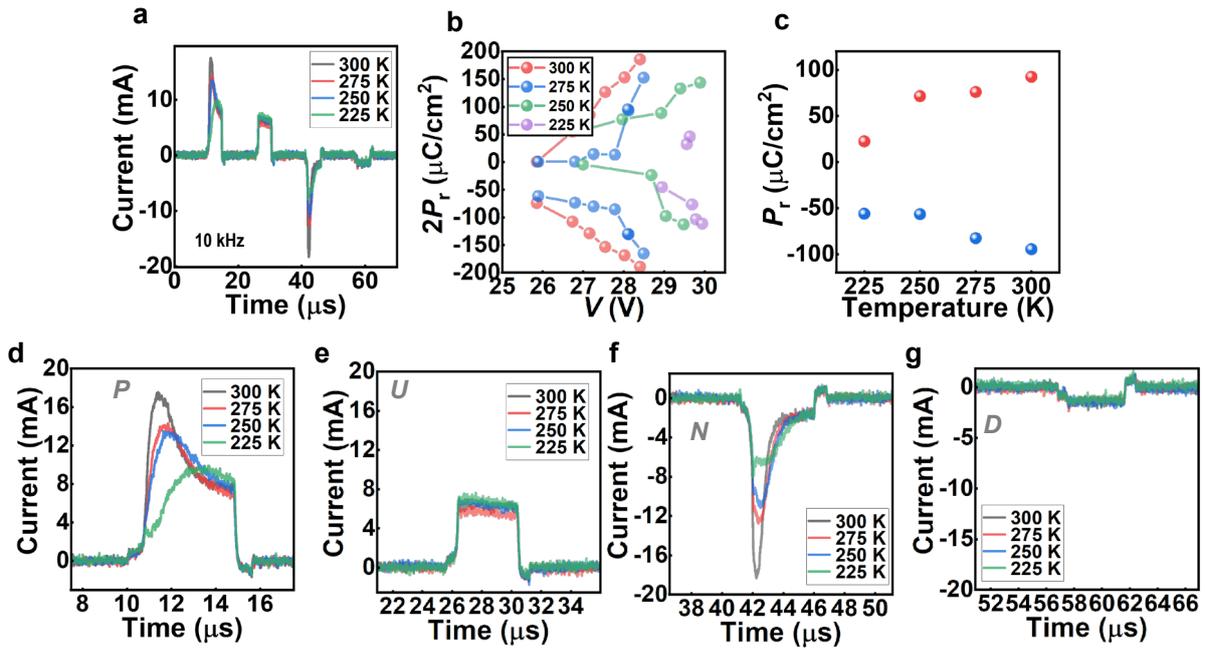

**Figure S5. PUND measurements of 45 nm AlScN grown with low $N_2$ flow.** (**a**) PUND results of the 45nm AlScN capacitors at 225-300K. (**b**) PUND-derived $2P_r$ values depending on the applied voltage. (**c**) PUND-derived $P_r$ values depending on the Temperature. (**d-g**) P, U, N, D pulse response.

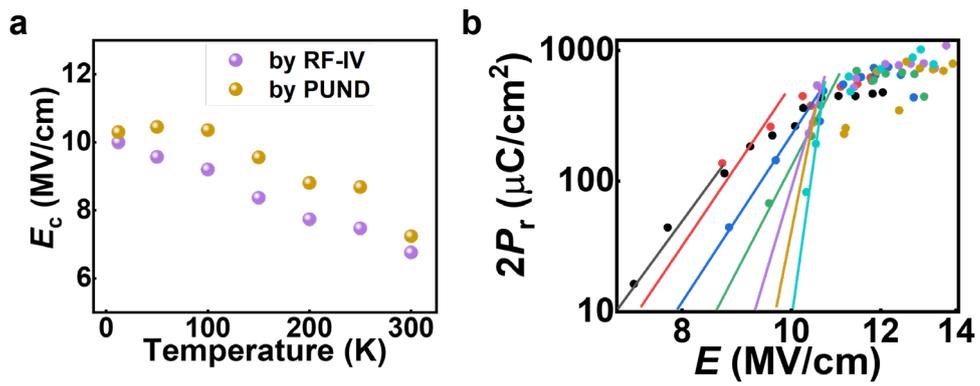

**Figure S6. PUND measurements of 10-nm-thick AlScN ferroelectric capacitors with via-connected devices.** (**a**) Comparison of the coercive field ($E_c$) extracted from both hysteresis and PUND measurements. (**b**) Illustration of the procedure used to determine the saturated $2P_r$ in Fig. 4b.

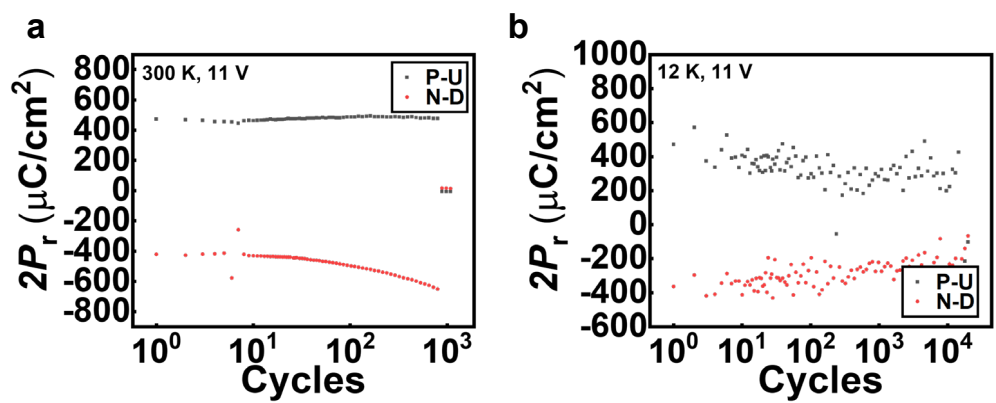

**Figure S7. Endurance test of 10 nm AlScN at 300 K and 12 K for P-U and N-D 2*Pr*.** (**a**) Endurance test at 300 K. (**b**) Endurance test at 12 K.

Table S1. Comparison of $2P_r$ and $E_c$ of AlScN with other well-known ferroelectric materials at room temperature and low temperatures.

| Ferroelectric material | Thickness and method | $2P_r$ @RT (µC/cm$^2$) | $2P_r$ @Low T (µC/cm$^2$) | $E_c$ @RT (MV/cm) | $E_c$ @Low T (MV/cm) | Reference |
|---|---|---|---|---|---|---|
| **Al$_{0.68}$Sc$_{0.32}$N** | **10 nm Sputtering** | **~250** | **~700 @12K** | **~6** | **~10 @12K** | This work |
| Hf$_{0.5}$Zr$_{0.5}$O$_2$ | 20nm ALD | 23 | 20 @100K | ~0.7 | ~0.8 @100K | 1 |
| Hf$_{0.5}$Zr$_{0.5}$O$_2$ | 10nm ALD | 60 | 150 @4K | 2.4 | 3 @4K | 2 |
| Hf$_{0.95}$Si$_{0.05}$O$_2$ | 10nm ALD | 76 | 94 @4K | 1.1 | 1.38 @100K | 3 |
| PbZr$_{0.5}$Ti$_{0.5}$O$_3$ | 300nm sol-gel process | 68 | 55 @4K | 0.11 | 0.36 @4K | 4 |
| Pb$_{0.990}$(ZrTi$_{0.47}$(Sb$_{0.67}$Nb$_{0.33}$)$_{0.08}$)O$_3$ | 1mm ground | 56 | 62 @123K | 0.01 | 0.014 @123K | 5 |
| SrBi$_2$Ta$_2$O$_9$ | 200nm PLD | 32 | ~1 @10K | 0.035 | - | 6 |

## Supplementary References